\documentclass[epjST]{svjour}
\usepackage{doi}
\usepackage{sidecap}
\usepackage{hyperref}
\hypersetup{
 colorlinks=true,
 linkcolor=blue,
 filecolor=magenta, 
 urlcolor=cyan,
 pdftitle={Nonequilibrium Higgs},
 }
\usepackage{amsmath}
\usepackage{slashed}
\usepackage{graphicx}
\usepackage{bm}
\usepackage{float}
 \usepackage{amssymb}
\usepackage[usenames,dvipsnames]{xcolor}

\usepackage{graphicx,wrapfig,lipsum}
\usepackage{subeqnarray}
\usepackage{xcolor}
\usepackage{eqnarray,amsmath}
\newcommand{\be}{\begin{equation}}
\newcommand{\beq}{\begin{equation}}
\newcommand{\ee}{\end{equation}}
\newcommand{\eeq}{\end{equation}}
\newcommand{\ba}{\begin{eqnarray}}
\newcommand{\ea}{\end{eqnarray}}
\newcommand{\ban}{\begin{eqnarray*}}
\newcommand{\ean}{\end{eqnarray*}}

\newcommand{\req}[1]{Eq.\,({\ref{#1}})}
\newcommand{\rf}[1]{Fig.\,{\ref{#1}}}
\newcommand{\rsec}[1]{Section\,{\ref{#1}}}
\newcommand{\orcJ}{0000-0001-8217-1484}
\newcommand{\orcC}{0000-0001-5038-8427}
\newcommand{\orcS}{0009-0009-9589-5532}
\newcommand{\orcidicon}{\includegraphics[width=0.32cm]{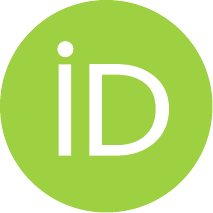}}
\newcommand{\orc}[1]{\href{https://orcid.org/#1}{\orcidicon}}
\begin{document} 
\title{Higgs Thermal Nonequilibrium in
Primordial QGP
}
\author{Cheng Tao Yang\orc{\orcC}, Shelbi Foster\orc{\orcS}, Johann Rafelski\orc{\orcJ}
}
\institute{Department of Physics, The University of Arizona, Tucson, Arizona 85721, USA}
\date{6 December 2024}

\abstract{In this work we investigate the chemical and kinetic nonequilibrium dynamics of the Higgs boson during the primordial Universe Quark-Gluon Plasma (QGP) epoch $130\mathrm{\,GeV}>T>10\mathrm{\,GeV}$. We show that the Higgs bosons is always out of chemical abundance equilibrium with a fugacity $\Upsilon_h = 0.69$ due to virtual decay channels. Additionally, Higgs momentum distribution is found to be ``cold'' for $T<25$\,GeV, since the scattering rate drops below the production rate.
}
\maketitle
\section{Introduction}
The Higgs particle is the second heaviest fundamental particle known today, yet it is relatively stable, with a life span about 500 times longer compared to the next heaviest gauge particles $W^\pm,Z^0$~\cite{ParticleDataGroup:2022pth}. Moreover, Higgs mass is below the threshold that allows decay into a pair of these massive heavy particles: The `minimal' Higgs coupling to massive particles weakens as their mass decreases. Therefore, both in the laboratory and in the primordial Universe, the production and decay processes can be different from each other as well as different from scattering: This situation is shown in \rf{HiggsDiagram_fig} where the Feynman diagrams for Higgs production (top line, (a) and (b)), and decay (top row, (c)) are shown. In bottom row we see (d) the dominant scattering processes in a thermal environment.

\begin{figure}
\begin{center}
\includegraphics[width=0.85\columnwidth]{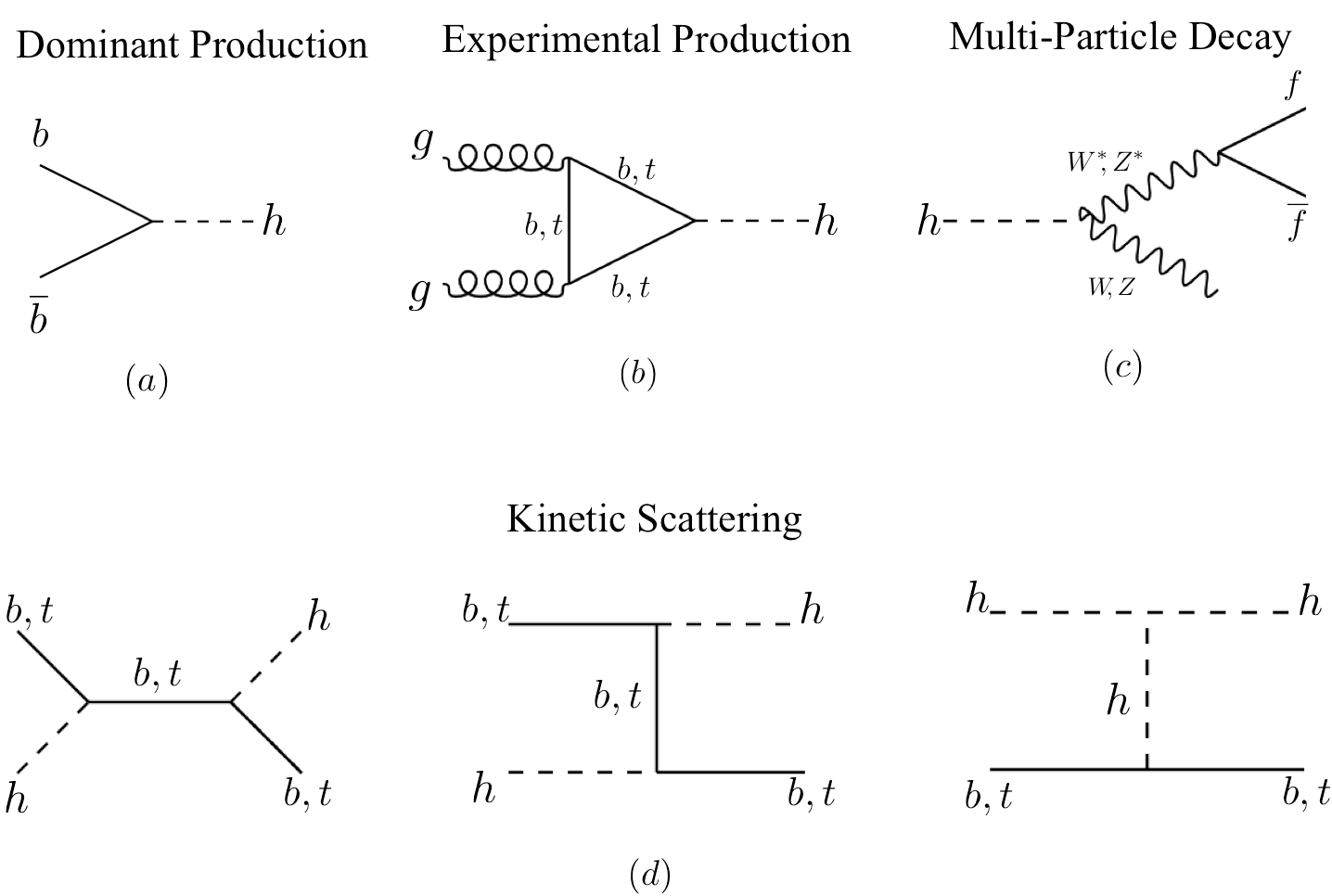}
\caption{The tree-level Feynman diagrams for Higgs production and scattering. $(a)$ The dominant Higgs production in QGP. $(b)$ The experimental production of Higgs. $(c)$ The multi-particle decay via virtual $W^\ast,Z^\ast$ bosons. $(d)$ The kinetic scattering between Higgs and $b,t$-quarks in QGP. }
\label{HiggsDiagram_fig}
\end{center}
\end{figure}

The two-to-one Higgs particle production is dominant. The decay includes relevant one-to-three (and more) real particles; the virtual gauge bosons $W^\ast,\,Z^\ast$ cannot persist~\cite{Glover:1988fn}, see (c) process in \rf{HiggsDiagram_fig}. Since the virtual particle decay amplitudes sum in probability to unity, the decay rate is not suppressed by the weak interaction that governs the secondary decay processes. However, each inverse reaction is suppressed by weak interactions as each $3\to 1$ channel is incoherent, involving `observed' real on mass shell fermion pairs. For this reason the decays which involve virtual particles near to their mass shell are in general not in the detailed balance condition~\cite{Vereshchagin:2017}. The magnitude of this defect will be characterized for the Higgs particle in this work. This leads to persistent small chemical non-equilibrium of the Higgs abundance. Let us repeat: This very special situation arises for the Higgs particle since at least one of the two heavy decay particle to which Higgs is more strongly minimally coupled has to be virtual considering the below pair threshold value of the Higgs mass. 

To add even more complexity, the momentum exchanging scattering processes in primordial Quark-Gluon Plasma (QGP) involve as usual two-on-two particle scattering. Three typical processes are seen in \rf{HiggsDiagram_fig} in bottom line (d)). These diagrams after $s\to t$ Mandelstam variable crossing do not relevantly contribute in Higgs particle abundance. In this work, we show that as a consequence of this situation -- {\it i.e.\/} the lack of connection of scattering to production and decay processes the Higgs does not achieve kinetic equilibrium below a certain temperature. This is true even if we assume that Higgs is in abundance (chemical) equilibrium. To best of our knowledge this is the only recognized system allowing the existence of kinetic non-equilibrium in the presence of particle abundance (chemical) equilibrium within relativistic particle and plasma context.

Clearly the Higgs content and kinetic momentum distribution in the primordial hot Universe is a topic requiring further detailed study within full kinetic theory context, which present work begins. One could argue this would be done just to satisfy our curiosity. However, the Higgs boson is a cornerstone of the Standard Model of particle physics, and holds significant importance in understanding the fundamental forces and particles that govern our Universe. Indeed there maybe more immediate relevance to this topic. As we argue in this work, the Higgs particle is out of thermal equilibrium for a long time after electroweak phase transition. This means that the electro-weak phase transition did not end rapidly. This could mean that baryogenesis possible due to nonequilibrium processes at the elctro-weak phase transition could continue for much longer in the kinetically evolving Universe. For further consideration of baryogenesis we defer here to the excellent reviews of Canetti, Drewes, and Shaposhnikov~\cite{Canetti:2012zc}, and Morrissey and Ramsey-Musolf~\cite{Morrissey:2012db}; we focus our attention on the understanding of the Higgs non-equilibrium in QGP phase of the Universe. 

One can wonder how the general assumption of total equilibrium in the primordial Universe has come to be. The reasoning is based on the magnitude of the characteristic Universe expansion time $\tau_\mathrm{U}=1/H$, considering the inverse of the Hubble parameter $H=\dot a(t)/a(t)$. Here $a(t)$ is the expansion scale of the Universe entering the cosmological Friedman metric. Using the Friedmann-Lemaitre-Robertson-Walker (FLRW) cosmology model dominated by matter and radiation, we evaluate the well known relation with the material energy density $\rho_i$
\begin{align}
H^2=\frac{8\pi G}{3}\left(\rho_\gamma+\rho_{\mathrm{lepton}}+\rho_{\mathrm{quark}}+\rho_{g,{W^\pm},{Z^0}}\right),
\label{H2Friedman}
\end{align} 
where $G$ is the Newtonian constant of gravitation. In the primordial Universe within a temperature range $130\, \mathrm{GeV}>T>0.15\, \mathrm{GeV}$ we have in the deconfined QGP the following particles: photons, $8_{color}$-gluons, $W^\pm$, $Z^0$, three generations of $3_{color}$-quarks, and leptons. The Einstein cosmological constant-style dark energy is irrelevant in this primordial epoch as is dark matter in any form compatible with present day dynamic Universe: both components are visible today due to extreme dilution of the Universe in subsequent expansion~\cite{Rafelski:2024fej}. The primordial QGP present in the early Universe during a temperature range of $130\,\mathrm{GeV} > T > 0.15\,\mathrm{GeV}$, is characterized by time scales $\tau_\mathrm{U}=10^{-9}$\,s and $\tau_\mathrm{U}=10^{-5}$\,s, respectively. 

During this primordial epoch, the Universe was dominated by strongly interacting particles: quarks and gluons. The magnitude of the microscopic strong QCD force reaction rates have been studied in depth~\cite{Letessier:2002ony}: The lifespan of laboratory QGP is of comparable magnitude as the relevant QCD reaction rates. The time scale governing experimental QGP created in relativistic heavy-ion collisions is of the order of magnitude $10^{-22}$\,s and shorter. This value follows from size of the system formed in laboratory. These laboratory scales are at least 13 orders of magnitude faster and can not compete with expansion of the Universe after electro-weak phase transition, even though the farther back in time we explore, the faster the Universe expansion dynamics are. Therefore the conclusion is very tempting that all particles in the Universe are as much thermally equilibrated as is the strongly interacting components. As we will show, this argument certainly fails for the minimally coupled Higgs.

Let us briefly digress to gain better understanding of what types of non-equilibrium conditions can arise. Thermal equilibrium requires both chemical equilibrium in which particle abundances are at a maximum (`black body' yield, when speaking about photons) and kinetic equilibrium in which energy has been shared and distributed to maximize entropy given the number of particles, thus according to the thermal quantum Fermi, Bose distributions. The exploration of heavy strange flavor particle abundance in relativistic heavy ion collisions has driven historically the more complete comprehension of these two different elements needed to achieve total thermal equilibrium. Only once mass threshold is overcome by sufficiently high kinetic energy (temperature $T$), massive particles can be produced abundantly and full thermal equilibrium can be achieved.

The idea that chemical equilibrium could not be achieved in presence of slow particle production processes in relativistic heavy ion collisions was first proposed in~\cite{Biro:1981zi}, and further developed showing time dependent approach to equilibrium of strangeness in QGP in~\cite{Rafelski:1982pu}. For a review of the early developments and study of the approach to equilibrium see~\cite{Koch:1986ud}, where also in chapter 6.3 the time dependent strangness abundance fugacity $\gamma$ was proposed: Chemical non-equilibrium can be described by introducing the pair abundance fugacity parameter now in general called $\Upsilon$ in the Fermi/Bose distribution~\cite{Letessier:1993qa}: 
\begin{align}
f_{F/B}(\Upsilon_i,p_i)=\frac{1}{\Upsilon^{-1}_i\exp{\left[E(p_i)/T\right]}\pm1}
\,,
\label{quantumDist}
\end{align}
where the plus sign applies for fermions, and the minus sign for bosons. The condition $\Upsilon=1$ indicates chemical equilibrium, while the deviations $\Upsilon\neq 1$ implies a departure from chemical equilibrium. Kinetic equilibrium is usually, {\it but not necessarily}, established faster through scattering processes, with little impact on chemical equilibrium.

Our manuscript is organized as follows: we describe in the following \rsec{sec:entro} the ideas which allow us to relate time to temperature so that we can explore properties of the Universe as function of time and compare with time dependent processes involving the Higgs particle. In \rsec{sec:Hpart}, we relate the stationary abundance of heavy particles in primordial QGP to ambient temperature. In \rsec{sec:Anoneq}, we obtain the rates of production of the Higgs particle which we find to be overall faster than the expansion of the Universe. However, detailed balance is broken since decay into two Gauge bosons of which one is virtual does not have a back reaction; the inverse $3\to 1$ process is of higher order in weak interaction and is suppressed by relevant coupling constants $g^2, g^{\prime\,2}$. This assures that Higgs abundance is always out of chemical equilibrium. In \rsec{sec:Snoneq} we obtain the scattering rates of Higgs particle in QGP and show these are slower compared to Higgs production below $T=25$\,GeV. This means that the distribution in momentum is a result of production process and not scattering, at low $T$ this implies that Higgs momentum distribution is also out of equilibrium. We quantify, and discuss further these findings in our discussion, \rsec{sec:Disc}.

\section{Entropy and baryon content of the Universe}\label{sec:entro}
An important assumption allowing us to explore the primordial Universe evolution is that following the era of matter genesis, both baryon and entropy density are conserved in the comoving volume--comoving in the sense that as the scale parameter $a(t)$ increases we look at a volume scaled up with $a(t)^3$. Therefore the ratio of baryon number density to visible matter entropy density remains constant throughout the evolution of universe as long as there is no significant entropy production. We have
\begin{align}\label{BaryonEntropyRatio}
\frac{n_B-n_{\overline{B}}}{s_{\mathrm{QGP}}}= \left.\frac{n_B-n_{\overline{B}}}{ s_{\gamma,\nu}}\right|_{t_0}=\mathrm{Const.}=\left(\frac{n_B-n_{\overline{B}}}{n_\gamma}\right)\left(\frac{n_\gamma}{s_\gamma+s_\nu}\right)_{\!t_0}\!\!\!=(8.69\pm0.05)\times10^{-11},
\;
\end{align}
The subscript $t_0$ denotes the present day condition, allowing us to fix the values that also apply to the primordial Universe in the absence of baryogenesis and entropy injection. Current observation gives the present baryon-to-photon ratio~\cite{ParticleDataGroup:2022pth} $5.8 \times 10^{-10} \leqslant(n_B-n_{\overline{B}})/n_\gamma\leqslant6.5\times10^{-10}$. This small value quantifies the matter-antimatter asymmetry in the present day universe and allows the determination of the present value of baryon per entropy ratio. The value ${(n_B-n_{\overline{B}})}/{n_\gamma}=(6.12\pm0.04)\times10^{-10}$ is used in the calculation.

We have considered the Universe today to be containing photons and free-streaming massless neutrinos~\cite{Birrell:2012gg}, and $s_\gamma$ and $s_\nu$ are the entropy densities for photons and neutrinos, respectively. We have
\begin{align}
 \frac{s_\nu}{s_\gamma}=\frac{7}{8}\,\frac{g_\nu}{g_\gamma}\left(\frac{T_\nu}{T_\gamma}\right)^3\,,\qquad\frac{T_\nu}{T_\gamma}=\left(\frac{4}{11}\right)^{1/3}
 \,,
\end{align}
and the entropy-per-particle\index{entropy!per particle} for massless bosons and fermions are given by~\cite{Fromerth:2012fe}
\begin{align}
s/n|_\mathrm{boson}\approx 3.60\,,\qquad
s/n|_\mathrm{fermion}\approx 4.20\,.
\end{align}

The evaluation of entropy of free-streaming fluid in terms of effectively massless $m\,a_f/a(t)$ free-streaming particles (neutrinos) needs further consideration, as does the free-streaming particles entropy definition. We will return to these important questions in the near future.

The entropy density in QGP can be written employing the effective number of `entropy' degrees of freedom $g^s_\ast$ 
\begin{align}
 &s_{\mathrm{QGP}}=\frac{2\pi^2}{45}g^s_\ast T_\gamma^3\,,\qquad 
g^s_\ast=\!\!\sum_{i=\mathrm{bosons}}\!\!g_i\left({\frac{T_i}{T_\gamma}}\right)^3B\left(\frac{m_i}{T_i}\right)+\frac{7}{8}\sum_{i=\mathrm{fermions}}\!\!g_i\left({\frac{T_i}{T_\gamma}}\right)^3F\left(\frac{m_i}{T_i}\right).
\end{align}
The mass dependent functions $B(m_i/T)$ and $F(m_i/T)$ are 
\begin{align}
&B\left(\frac{m_i}{T}\right)=\frac{15}{4\pi^4}\int^\infty_{m_i/T}\,dx\frac{\sqrt{x^2-\left({m_i}/{T}\right)^2}\left[4x^2-\left({m_i}/{T}\right)^2\right]}{\Upsilon^{-1}_ie^x-1}\,,\\
&F\left(\frac{m_i}{T}\right)=\frac{30}{7\pi^4}\int^\infty_{m_i/T}\,dx\frac{\sqrt{x^2-\left({m_i}/{T}\right)^2}\left[4x^2-\left({m_i}/{T}\right)^2\right]}{\Upsilon^{-1}_ie^x+1}\,,
\end{align}
where $\Upsilon_i$ is the fugacity parameter for a given particle. 

When $T$ decreases below the mass of the particle ($T\ll m_i$) and becomes non-relativistic, the functions $B(m_i/T)$ and $F(m_i/T)$ go to zero, which implies that the contribution of the non-relativistic species to $g^s_\ast$ is negligible. The dominant factor $T^3$ cancels when computing the ratio of photons per entropy; therefore, the ratio of baryon number density to visible matter entropy density remains constant throughout the evolution of the Universe.

To determine the relation between time and temperature as the Universe evolves we first consider comoving entropy conservation\index{entropy!conservation}, 
\begin{align}
S=\sigma V\propto g^s_\ast T^3a^3=\mathrm{constant},
\end{align}
where $g^s_\ast$ is the entropy degree of freedom and $a$ is the scale factor. Differentiating the entropy with respect to time $t$ we obtain
\begin{align}
\left[\frac{\dot{T}}{g^s_\ast}\frac{dg^s_\ast}{dT}+3\frac{\dot{T}}{T}+3\frac{\dot{a}}{a}\right]g^s_\ast T^3a^3=0,\qquad \dot{T}=\frac{dT}{dt}.
\end{align}
The square bracket has to vanish. Solving for $\dot T $ we obtain
\begin{align}
\frac{dT}{dt}=-\frac{HT}{1+\frac{T}{3g^s_\ast}\frac{d\,g^s_\ast}{dT}}\,.
\end{align}
In our approach this relation is a smooth function even when the number of degrees of freedom changes as we allow for finite mass of particles. 

\section{Heavy particles in the primordial QGP}\label{sec:Hpart}
Considering minimal coupling and the structure of Higgs effective action, the key heavy particles (aside from itself=self interaction) that allow Higgs to equilibrate thermally are the top quark $t$, and the $Z^0, W^\pm$ weak interaction mediating vector mesons. A convenient way to study their abundance is to compare to the baryon yield residing in the quark-antiquark (nano-sized) asymmetry. 

The thermal equilibrium number density of heavy particles with mass $m\gg T$ can be well described by the Boltzmann expansion of the Fermi distribution function, giving
\begin{align}\label{BoltzN}
n_{i}\!=\!\frac{g_{i}T^3}{2\pi^2}\sum_{n=1}^{\infty}\frac{(-1)^{n+1}\Upsilon_i^n}{n^4}\left(\frac{n\,m_{i}}{T}\right)^{\!2}\!K_2\left(\frac{n\,m_{i}}{T}\right),
\end{align} 
where $i=t,W,Z,h$; $\Upsilon$ is the fugacity parameter, and $K_2$ is the modified Bessel function of the second kind of integer order $2$. In the temperature range we consider here, we have $m_i\gg T$ which allows us to consider the Boltzmann limit and keep only the first term $n=1$ in the expansion. In this scenario,
the number density of the heavy particle becomes
\begin{align}\label{Density}
&n_{i}=\Upsilon_i\,n^{th_i},\quad n^{th}_i=\frac{g_i}{2\pi^2}T^3\left(\frac{m_i}{T}\right)^2 K_2(m_i/T)
\end{align}
where $n^{th}_i$ corresponds to the thermal equilibrium number density of given particle $i$.

\begin{SCfigure}
\includegraphics[width=0.65\columnwidth]{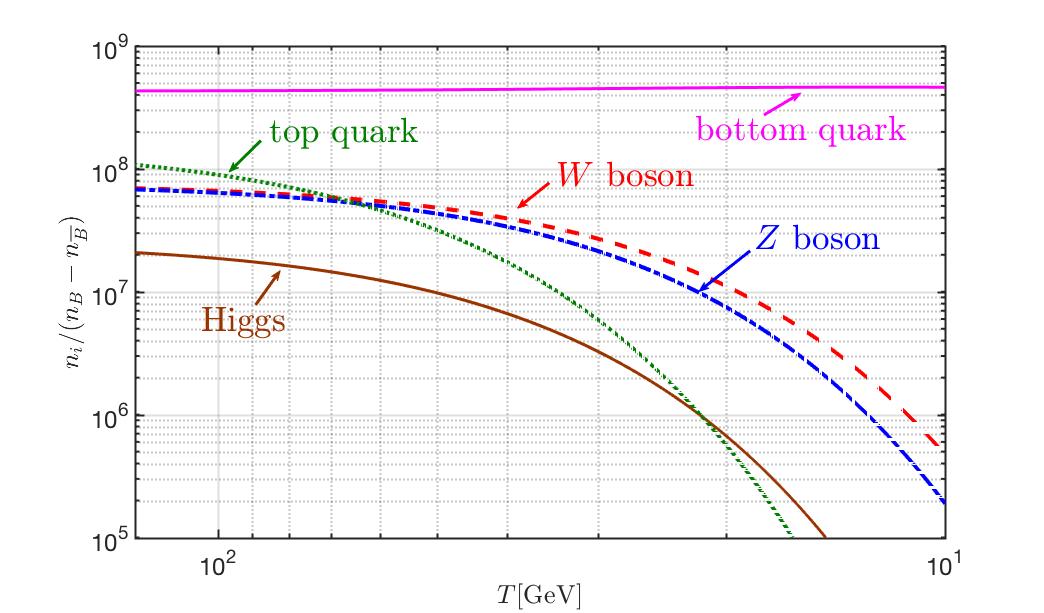}
\caption{The thermal equilibrium ($\Upsilon_i=1$) density ratio between heavy particle and baryon asymmetry as a function of temperature. This result shows that the heavy particle density is significantly larger than
the baryon number density.}
\label{HiggsDensity_fig}
\end{SCfigure}

Using a constant baryon-per-entropy ratio, the density between Higgs and baryon asymmetry ($u,d$ quark-antiquark asymmetry) can be written as
\begin{align}
\frac{n_h}{(n_B-n_{\bar{B}})}=\frac{n_{h}}{s_{\mathrm{QGP}}}\,\left(\frac{s_{\mathrm{QGP}}}{n_B-n_{\bar{B}}}\right)=
\frac{n_{h}}{s_{\mathrm{QGP}}}\left(\frac{s_{\gamma,\nu}}{n_B-n_{\bar{B}}}\right)_{\!t_0},
\end{align}
Where the subscript $t_0$ denotes the present day condition and $s_{QGP}$ is the total entropy density in QGP. In \rf{HiggsDensity_fig}, we show the thermal equilibrium ($\Upsilon_i=1$) number density ratio between heavy particles and baryon asymmetry. We show that the heavy particle density is significantly larger than the baryon number density, even at temperature $T=10$\,GeV, at which the ratio is about $10^5$.

\section{Chemical Nonequilibrium}\label{sec:Anoneq}
The decay of the Higgs boson into two real $W$ or $Z$ boson pairs is forbidden because the mass of the Higgs ($m_H=124$\,GeV) is less than double that of the $W$ ($ m=80.4$\,GeV) or $Z$ ($m_Z=91.2$\,GeV). Even so, the three main decay modes of Higgs are $h\rightarrow b+\bar b$, $h\rightarrow W+W^*$, and $h\rightarrow Z+Z^*$ where $W^\ast,Z^\ast$ represent the virtual bosons~\cite{Glover:1988fn}. The total decay width of the Higgs is $\Gamma_\mathrm{decay}=3.7$ MeV, and The branching ratios are $53\pm 8\%$ for the bottom decay channel, $25.7\pm 2.5\%$ for the W decay channel, and $2.8\pm 0.3\%$ the $Z$ decay channel~\cite{ParticleDataGroup:2022pth}. Because of the large branching ratio of bottom decay channel, the dominant production of the Higgs boson in QGP is the bottom fusion reaction: 
\begin{align}
b+\overline{b}\longleftrightarrow h,\qquad B_b=0.53
\end{align}
which is the inverse decay process of $H\to b+\overline{b}$, and $B_b$ is the branching ratio for bottom decay channel. On the other hand, Higgs abundance disappears via the $W,Z$ decay channel as follows:
\begin{align}
h\longrightarrow WW^\ast, ZZ^\ast\longrightarrow\mathrm{anything},\qquad B_{W,Z}=0.285,
\end{align}
where $B_{W,Z}$ is the branching ratio for Higgs decay into $W,Z$ bosons.

After being produced through Higgs decay, the virtual bosons $W^\ast$ and $Z^\ast$ immediately decay into other particles, as they must disappear the rate is not hindered by existence of weak interaction process driving the decay - that is why the Higgs decays so abundantly into the virtual pairs. Another consequence of the large masses of $W$ and $Z$ is that the fusion reaction $WW, ZZ\rightarrow H$ is kinematically forbidden. Considering two fermions producing the off-mass shell gauge meson we obtain a possible inverse process which is suppressed in the evaluation by the weak interaction that allows two fermions to combine with one real gauge meson into the Higgs. Consequently, the inverse of multi-particle decay contributes only about $1\%$ to Higgs abundance.

The reaction rate per time per volume for the inverse decay reaction $1+2\to3$ has been thoroughly studied in the paper~\cite{Kuznetsova:2010pi,Kuznetsova:2008jt}. The thermal rate per unit time and volume can be expressed as
\begin{align}
R_{12\to 3}=\frac{g_3}{(2\pi)^2}\,\frac{m_3}{\tau^0_3}\,\int^\infty_0\frac{p^2_3dp_3}{E_3}\frac{e^{E_3/T}}{e^{E_3/T}\pm1}\Phi(p_3)
\end{align}
where $\tau_3^0$ is the vacuum lifespan of heavy particle 3. In the temperature range of our interests, we have $m_H\gg T$, and the nonrelativistic Boltzmann approximation is suitable for studying massive particles. The function $\Phi(p_3)$ in the nonrelativistic limit is given by 
\begin{align}
\Phi(p_3\to0)=2\frac{1}{(e^{E_1/T}\pm1)(e^{E_2/T}\pm1)}.
\end{align}
In Boltzmann limit, the thermal decay rate per unit volume and time for heavy particle becomes
\begin{align}
R_{12\to 3}=\frac{g_3}{2\pi^2}\left(\frac{T^3}{\tau_3^0}\right)\left(\frac{m_3}{T}\right)^2K_1(m_3/T)\;,\
\end{align}
where $K_1$ is the modified Bessel function of the second kind of integer order 1. It is convenient to define the fusion rate for the process $1+2\to3$ as follows
\begin{align}
\Gamma_{12\rightarrow 3}\equiv\frac{R_{12\rightarrow 3}}{n^{th}_{3}}
\end{align}
In \rf{Fusion_fig}, we present the Higgs production rate from possible fermion fusion processes as well as the Hubble parameter as functions of temperature. This result shows that the dominant contribution to the Higgs boson production comes from the bottom quark fusion process. Furthermore, the fusion rates are significantly larger than the Hubble parameter. It implies that the expansion of Universe has a negligible effect on the abundance of Higgs bosons because of its short vacuum lifespan.

\begin{SCfigure}
\includegraphics[width=0.65\columnwidth]{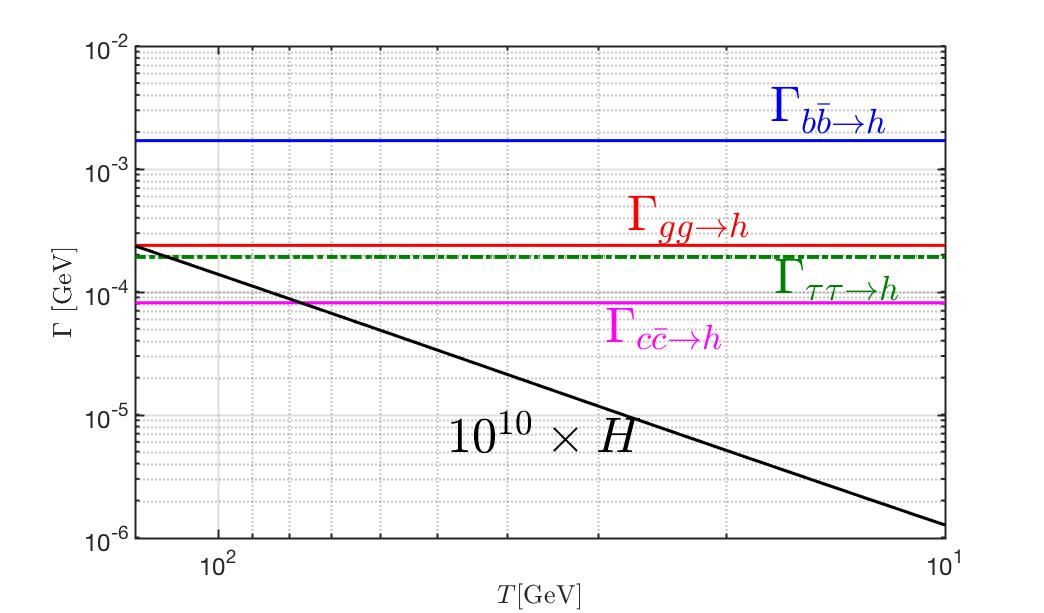}
\caption{The Higgs fusion rates as a function of temperature compare to the Hubble parameter (scaled by a factor of $10^{10}$). This result shows that the dominant production of the Higgs boson is the bottom fusion reaction, and the fusion rates are significantly larger than the Hubble parameter.}
\label{Fusion_fig}
\end{SCfigure}

Considering the production and decay reaction processes for the Higgs in QGP, the population equation that describes the rate of change in the number of Higgs particles per unit volume is given by
\begin{align}
\label{Higgs_eq}
\frac{1}{V}\frac{dN_h}{dt}
=\sum_{i=b,c,g,\tau}(\Upsilon_i-\Upsilon_h)R_{i\overline{i}\to h}-\Upsilon_h R_{h\rightarrow WW^\ast,ZZ^\ast},
\end{align}
where $\Upsilon_h$ is the Higgs fugacity parameter and $\Upsilon_i$ is the fugacity of the particle species $i$. Due to existence of the decay channel into virtual particles $Z^\ast, W^\ast$ the last term in~\req{Higgs_eq} is not balanced by an associated back-reaction term: The two body decay produces virtual bosons $W^\ast$ and $Z^\ast$ which materialize decaying into lighter on-mass real particles with unit probability, the total decay rate involves just one weak interaction coupling. On the other hand the back reaction requires collision of at least three on-mass shell particles, with higher 4, 5, 6, particle collision channels also contributing. Such back reaction channels require additional multiple powers of coupling parameters which for the electro-weak reactions are suppressing the inverse processes in a very significant manner. Therefore, we have omitted the back-reaction process entirely. Such non-equilibrium generating imbalance is unique to the Higgs boson since its mass is below but near to the $WW$ and $ZZ$ production threshold, and the coupling to these particles is very strong.

We consider quarks, gluons, and leptons in abundance equilibrium during the QGP epoch with fugacity $\Upsilon_i=1$. Then the Higgs population equation becomes
\begin{align}
\frac{1}{V}\frac{dN_h}{dt}
=(1-\Upsilon_h)R_\mathrm{fusion}-\Upsilon_h R_{h\rightarrow WW^\ast,ZZ^\ast},\qquad R_\mathrm{fusion}=\sum_{i=b,c,g,\tau}R_{i\overline{i}\to h}.
\end{align}
We aim to replace the variation of particle abundance seen on left hand side in \req{Higgs_eq} by the time variation of abundance fugacity $\Upsilon_H$. This substitution allows us to derive the dynamic equation for the fugacity parameter and enables us to study the fugacity as a function of time. 

Considering the expansion of the Universe, the left hand side in \req{Higgs_eq} can be written as
\begin{align}\label{number_dilution}
\frac{1}{V}\frac{dN_h}{dt}=\frac{1}{V}\frac{d(n_hV)}{dt}=\frac{dn_h}{d\Upsilon_h}\frac{d\Upsilon_h}{dt}+\frac{dn_h}{dT}\frac{dT}{dt}+3Hn_h,\;
\end{align}
where $H$ is the Hubble parameter and we use $d\ln(V)/dt=3H$ for the Universe expansion. Substituting \req{number_dilution} into \req{Higgs_eq} and dividing both sides of equation by $dn_H/{d\Upsilon_H}=n^{th}_H$, the fugacity equation becomes
\begin{align}\label{Fugacity_eq}
\frac{d\Upsilon_h}{dt}+\Upsilon_h &\left(\frac{dn_h^{th}/dT}{n_h^{th}}\frac{dT}{dt}+3H\right)=(1-\Upsilon_h)\frac{R_\mathrm{fusion}}{n_h^{th}} - \Upsilon_h\frac{R_{h\rightarrow WW^\ast,ZZ^\ast}}{n_h^{th}}.
\end{align}
Considering the nonrelativistic Boltzmann limit $m_H\gg T$ for the Higgs, we have 
\begin{align}
\frac{dn^{th}_h/dT}{n^{th}_h}\frac{dT}{dt}&=-\frac{H}{1+\frac{T}{3g^s_\ast}\frac{d\,g^s_\ast}{dT}}\left[3+\frac{m_h}{T}\frac{K_1(m_h/T)}{K_2(m_h/T)}\right]=-\frac{H}{1+\frac{T}{3g^s_\ast}\frac{d\,g^s_\ast}{dT}}\left[3+\frac{m_h}{T}\left(1-\frac{3}{2}\frac{T}{m_h}+\cdots\right)\right]\notag\\
&\approx-\frac{m_h}{T}\frac{H}{1+\frac{T}{3g^s_\ast}\frac{d\,g^s_\ast}{dT}}.
\end{align}
In \rf{Fusion_fig}, during the epoch of interest, the Hubble parameter is significantly smaller than the fusion/decay rates. Consequently, the terms related to the Hubble parameter can be neglected in \req{Fugacity_eq}, simplifying the fugacity equation for the Higgs to:
 \begin{align}\label{Eq:PopH}
\frac{d\Upsilon_h}{dt}\!\!
=(1-\Upsilon_h)\Gamma_\mathrm{fusion}-\Upsilon_h \Gamma_{h\rightarrow WW^\ast,ZZ^\ast},
 \end{align}
 where the total Higgs fusion rate and decay rate are given by:
\begin{align}
\Gamma_\mathrm{fusion}=\frac{R_{b\bar{b}\rightarrow h}+R_{c\bar{c}\rightarrow h}+R_{\tau\bar{\tau}\rightarrow h}+R_{gg\rightarrow h}}{n^{th}_h},\qquad
\Gamma_{h\rightarrow WW^\ast,ZZ^\ast}=\frac{R_{h\rightarrow WW^\ast,ZZ^\ast}}{n_h^{th}}.
\end{align}

Considering that at each given temperature we have the dynamic equilibrium condition, i.e. detailed balance between production and decay reactions that keep the abundance in nearly instantaneous stationary condition
\begin{align}
\frac{d\Upsilon_h}{dt}=0.
\end{align}
Under this condition, we solve the fugacity equation and obtain
\begin{align}
\Upsilon_h=\frac{\Gamma_\mathrm{fusion}}{\Gamma_\mathrm{fusion}+\Gamma_{H\to WW^\ast,ZZ^\ast}}=\frac{\Gamma_\mathrm{fusion}}{\Gamma_\mathrm{decay}}=0.69.
\end{align}
We have neglected the very weak temperature dependence of this non-equilibrium ratio arising in the stationery approximation $H(T)=Const.$ for the Universe. This is possible since the expansion rate is more than 10 order of magnitude slower compared to the microscopic reactions we consider in the entire temperature range $125\,\mathrm{GeV}>T>0.1\,\mathrm{GeV}$ in which the deconfined QGP state exists.
This result shows that Higgs exhibits nonequilibrium behavior with $\Upsilon_h=0.69$ in primordial QGP. This occurs due to the dynamic equilibrium condition (detailed balance) between production and decay reaction of Higgs.

\section{Kinetic Nonequilibrium}\label{sec:Snoneq}
Once the Higgs bosons are produced, kinetic scattering between the Higgs and quarks enables momentum exchange between them. In the primordial QGP, the primary interaction between quarks and the Higgs is the Compton-like scattering:
\begin{align}
b+h\longrightarrow {b}+{h},\qquad t+h\longrightarrow {t}+{h}.
\end{align}
Their lowest-order Feynman diagrams are shown in \rf{HiggsDiagram_fig}. The tree-level amplitudes for different channels are: 
\begin{align}
 &\mathcal{M}=\left(\frac{m_b}{v}\right)^2\frac{\slashed{p}_1+\slashed{p}_2+m_b}{(p_1+p_2)^2-m^2_b}\,\,u(p_1)\overline{u}(p_4),\\
 &\mathcal{M}=\left(\frac{m_b}{v}\right)^2\frac{\slashed{p}_1-\slashed{p}_3+m_b}{(p_1-p_3)^2-m^2_b}\,\,u(p_1)\overline{u}(p_4),\\
 &\mathcal{M}=\left(\frac{3m_bm_h^2}{v^2}\right)\frac{1}{(p_1-p_3)^2-m^2_h}\,\,u(p_2)\overline{u}(p_4),
\end{align}
where $v=246$\,GeV is the vacuum expectation value, and $u(p_i)$ represents the spinors for bottom quark.
In general, the thermal reaction rate per unit time and volume for a two-body scattering process can be written as~\cite{Letessier:2002ony}
\begin{align}
R_{12\rightarrow34}=\frac{g_1g_2}{32\pi^4}\frac{T}{1+I_{12}}\!\!\int^\infty_{s_{th}}\!\!\!\!ds\,\sigma(s)\frac{\lambda_2(s)}{\sqrt{s}}K_1(\sqrt{s}/T),\qquad
\lambda_2(s)\equiv\left[s-(m_1+m_2)^2\right]\left[s-(m_1-m_2)^2\right],
\end{align}
where the cross section $\sigma(s)$ can be obtained by integrating the transition amplitude of the lowest-order Feynman diagrams. We define the scattering rate for Higg-bottom/top scattering as follow:
\begin{align}
 {\Gamma_{hq\rightarrow hq}}\equiv\frac{R_{hq\rightarrow hq}}{n^{th}_h},\qquad q=b,t.
\end{align}
It is also convenient to define the total scattering rate for Higgs
\begin{align}
\Gamma_\mathrm{Scattering}=\frac{R_{hb\rightarrow hb}+R_{ht\rightarrow ht}}{n^{th}_h},
\end{align}
In \rf{Scattering_fig}, we plot the relevant scattering and fusion rates for the Higgs, along with the Hubble parameter, as functions of temperature. For kinetic scattering, the dominant process when $T>17$\,GeV is Compton scattering between the Higgs and top quarks. Below $T<17$\,GeV, the Higgs-bottom scattering becomes the dominant kinetic process. The total scattering rate intersects with the total fusion rate at temperature $T=25$\,GeV. Both fusion and scattering rates significantly exceed the Hubble parameter across the temperature range, indicating that the Universe's expansion does not affect Higgs-related reactions. 

\begin{SCfigure}
\includegraphics[width=0.6\columnwidth]{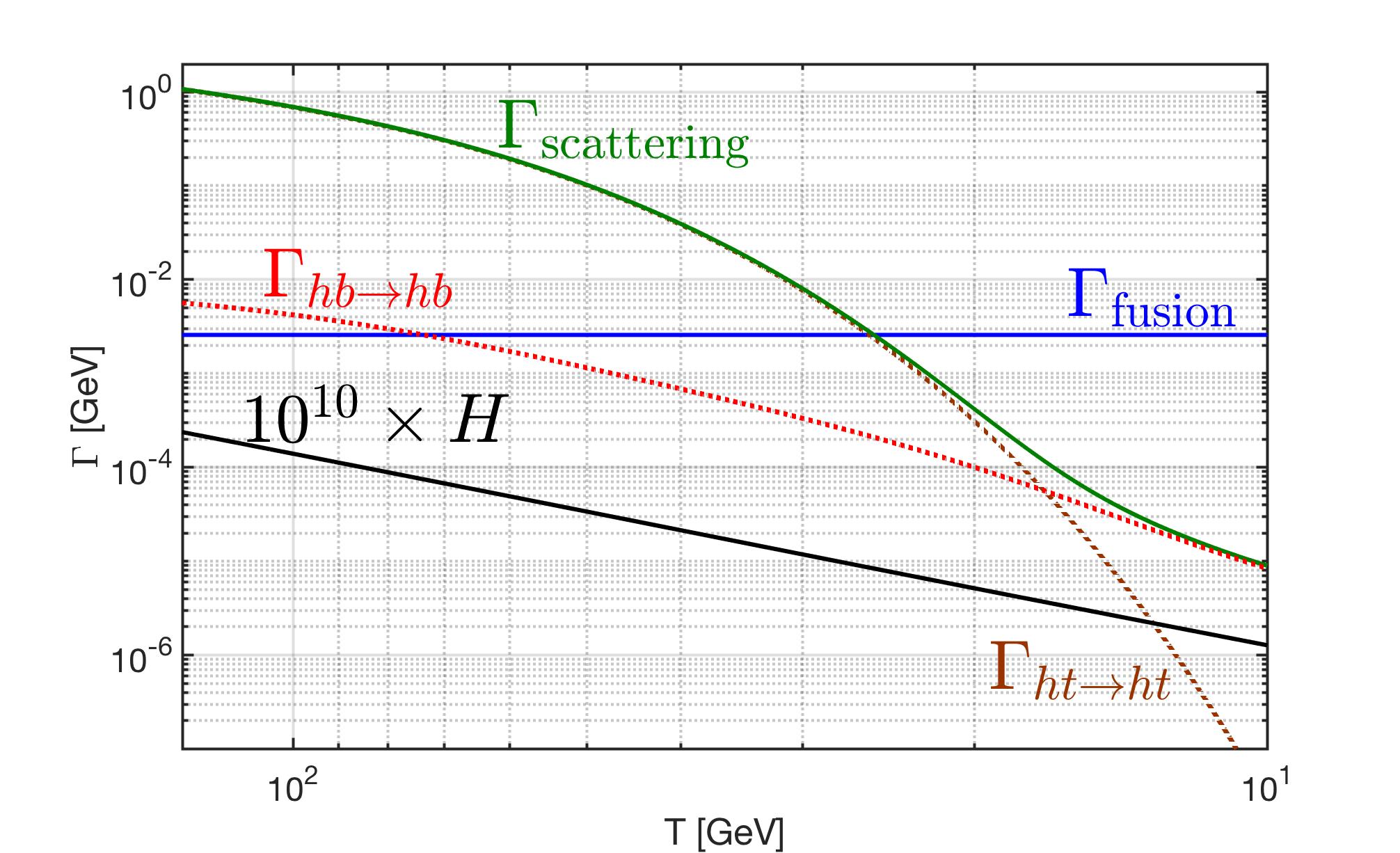}
\caption{The relevant scattering and fusion rates for the Higgs, along with the Hubble parameter (scaled by a factor of $10^{10}$), as functions of temperature. This result shows that op quark scattering dominates above $T=17$\,GeV, while bottom quark scattering takes over below this temperature. Both fusion and scattering rates significantly exceed the Hubble parameter across the temperature range.}
\label{Scattering_fig}
\end{SCfigure}

In \rf{Rate_Ratio_fig} we plot the ratio of Higgs scattering rate to fusion $\Gamma_\mathrm{scattering}/\Gamma_\mathrm{fusion}$ and decay $\Gamma_\mathrm{scattering}/\Gamma_\mathrm{decay}$ as functions of temperature. This result demonstrates that at temperatures $T>25$\,GeV, scattering dominates over both fusion to Higgs and Higgs decay, while for $T<25$\,GeV the scattering rate becomes smaller than the fusion rate. Therefore once `cold' Higgs bosons are produced via fusion process where particles in general are barely over mass threshold, there is insufficient kinetic scattering to exchange momentum with the background, preventing the thermalization of the produced Higgs and leaving it at in `cold' condition.

\begin{SCfigure}
\includegraphics[width=0.6\columnwidth]{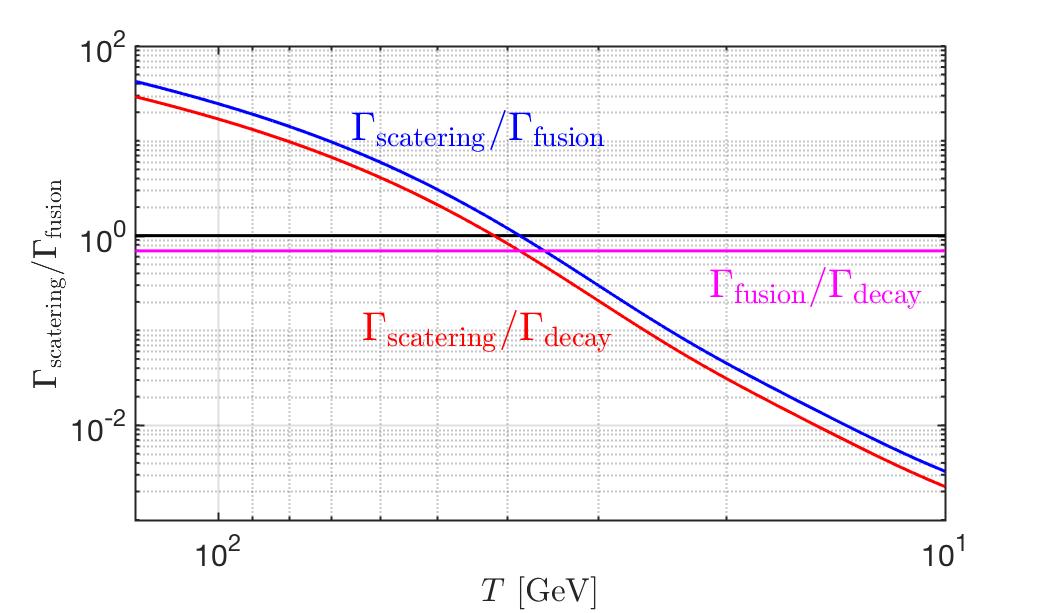}
\caption{The ratio of total Higgs scattering to fusion and decay rates as a function of temperature.The black solid line indicates a rate ratio of $1$. For temperatures $T>40$\,GeV, we have scattering larger than both fusion and decay $\Gamma_\mathrm{scattering}>\Gamma_\mathrm{fusion},\Gamma_\mathrm{decay}$. For temperatures $T<40$\,GeV, scattering becomes smaller than both fusion and decay $\Gamma_\mathrm{scattering}< \Gamma_\mathrm{fusion},\Gamma_\mathrm{decay}$.}
\label{Rate_Ratio_fig}
\end{SCfigure}

\section{Discussion}\label{sec:Disc}
In this work, we examined the chemical and kinetic equilibrium of the Higgs boson during the QGP epoch in the early Universe by analyzing the relevant reaction strengths. Our findings reveal that the Higgs boson remains out of both chemical and kinetic equilibrium.

Before turning to non-equilibrium processes, we have first evaluated the thermal equilibrium abundance of heavy particles in QGP. Figure~\ref{HiggsDensity_fig} shows the the number density ratio of heavy particles to net baryon density under assumption of total thermal equilibrium $(\Upsilon_i = 1)$. In the temperature range of interest above $T=10$\,GeV, the heavy particle density is significantly $10^5$ larger than the net baryon number density. 

In QGP, the dominant production mechanism for the Higgs boson is through the bottom-quark fusion. This process can be viewed as an inverse decay reaction ($1+2 \to 3$), where the natural decay properties of the Higgs dictate the strength of its inverse production as shown in \rf{Fusion_fig}. In contrast, the Higgs bosons are depleted primarily by decaying into the have gauge boson pairs, which due to mass threshold includes either a virtual $W^\ast$ or $Z^\ast$. Detailed balance is broken since a decay into two Gauge bosons of which one is virtual does not have a back reaction: the inverse process is of higher order in weak interaction and is suppressed by relevant coupling constants $g^2, g^{\prime\,2}$. Analyzing the dominant production and decay processes of the Higgs boson in the QGP, we solve the population equation for the Higgs and demonstrate its prolonged nonequilibrium behavior, characterized by a significant departure from thermal equilibrium with $\Upsilon_h=0.69$ as a consequence of the breach of detailed balance described: This chemical nonequilibrium state arises from the dynamic balance between Higgs production via bottom-quark fusion and its decay into vector bosons $W^\ast,Z^\ast$.

There is a second type of nonequilibrium which is due to ever weaker Higgs scattering rates in the QGP: scattering on light particles is negligible due to minimal coupling and the abundance of heavy particles decreases as temperature drops. This implies that Higgs momentum distribution is governed by the production process as the particle decays before experiencing a scattering. This distribution is not informed about ambient temperature especially at low $T$: The mass of a Higgs boson is much larger than that of a bottom quark pair producing it, which results in the production of cold Higgs through bottom quark fusion at sufficiently low $T$. In \rf{Scattering_fig} we present the relevant scattering and fusion rates for the Higgs, and show that that at $T<25$\,GeV the scattering rate is smaller compared to the fusion rate. This occurs because the kinetic scattering is a two-to-two particle reaction ($1+2\to 3+4$), which depends on the density of the particles. As the Universe expands and the temperature drops, the particle density decreases, reducing the probability of two-particle collisions and thereby decreasing the scattering rate. When the Universe expands and temperature cools down, the kinetic scattering slows down and is not able to keep up with fusion reaction, implying that the cold Higgs bosons produced through fusion struggle to thermalize with the surrounding background. This leads to the departure of the Higgs bosons from kinetic equilibrium in the early Universe.

In conclusion, our study provides valuable insights into the chemical and kinetic nonequilibrium behavior of Higgs during the primordial QGP epoch. These findings not only deepen the understanding of Higgs dynamics in QGP but also offer a potential framework for future research into early Universe nonequilibrium processes allowing presence of an arrow in time.\\

\textbf{Acknowledgment:} 
One of us (JR) thanks Tamas Bir\'o and the Wigner Hun-Ren Research Center for their kind hospitality in Budapest during the PP2024 conference, supported by NKFIH (Hungarian National Office for Research, Development and Innovation) under awards 2022-2.1.1-NL-2022-00002 and 2020-2.1.1-ED-2024-00314. This meeting and the related research report motivate the presentation of these recently obtained results. The authors and this work were not supported by any sponsor.



\bibliographystyle{sn-aps}


\end{document}